\documentstyle[twocolumn,aps,pra]{revtex}
\input epsf
\begin{document} 

\newcommand{\be}{\begin{equation}}
\font \cim=cmr10 scaled \magstep 4
\font \bev=cmr10 scaled \magstep 2
\newcommand{\prim}{\acute {}}
\newcommand{\ii}{\'\i}
\newcommand{\ee}{\end{equation}}
\newcommand{\td}{t_{\mathrm{d}}}
\newcommand{\tdiss}{t_{\mathrm{diss}}}

\draft

\title{
Rapid and slow decoherence in conjunction with dissipation in a system of two-level atoms
}

\author{
P\'eter F\"oldi\cite{PFemail}, 
Attila Czirj\'{a}k\cite{ACemail}, 
Mih\'{a}ly G. Benedict\cite{MGBemail}
}
\address{
Department of Theoretical Physics, University of Szeged, 
H-6720 Szeged, Tisza Lajos krt. 84-86, Hungary
}
\maketitle

\abstract{
We investigate the time evolution of a superposition
of macroscopically distinct quantum states 
in 
a system of two-level atoms 
interacting with 
a thermal environment of photon modes. 
We show that the atomic coherent states are robust against decoherence,
therefore we call their superpositions atomic Schr\"{o}dinger cat states.
The initial fast regime of the time evolution
is associated with the
process of decoherence, and it 
is directed towards the statistical  
mixture of the constituent coherent  states of the 
original state for most of the initial conditions. 
However, certain 
superpositions, called symmetric, exhibit
exceptionally slow decoherence. 
By introducing a new measure, we generalize the usual decoherence scheme 
regarding the evolution of the state to account also for the symmetric case. 
To stress the fact that the environment 
preserves symmetric superpositions 
much longer than the other ones,
we present Wigner function images of the 
decoherence of a suitably oriented four component cat state.
} 



\section {Introduction}

The apparent lack of a superposition  of macroscopically distinct quantum
states (Schr\"{o}dinger cats) has been an interesting and vivid problem since
Schr\"{o}dinger's famous paper \cite{sch}. 
A successful approach, initiated by Zeh \cite{zeh70} and developed by Zurek \cite{zurek981}, 
obtains the loss of quantum 
coherence as the consequence of the unavoidable interaction 
with the environment. Theoretical studies 
in this framework have investigated a variety of model systems usually coupled
to a collection of harmonic oscillators as an environment. 
Fundamental work has been done on this subject
in \cite{caldlegg983,unzur989,hupaz992,dekker981,sawalls985,diosi989,knight999}, 
for reviews see \cite{decref1,decref2}.
Important experiments have been carried out during the last years \cite{brune996,myatt000,Friedman}
and the possible production of long-lived quantum superpositions has gained 
wide attention because of their expected use in quantum computation \cite{benndivin000}. 

In the present paper we investigate a system which is a candidate for the 
experimental study of decoherence and possibly also for practical applications. 
The model 
to be described in detail in section \ref{model}, 
consists of  
several identical two-level atoms (the system) 
interacting with a  large number of photon 
modes in a thermal state (the environment). 
It has the advantage that it is simple to make the correct transition from a 
microscopic system to a macroscopic one by increasing the number of atoms.
We 
analyze the evolution of the reduced density 
matrix of the atomic system by the appropriate master equation 
\cite{bsh971a,agarwal974,haroche985}.
One of the main results of the present work is that we show how 
one and the same solution of this master equation 
describes both decoherence and dissipation.

By analytical short-time calculations we show in section \ref{robust} 
that the atomic coherent states \cite{acgt972} of our system are robust 
against 
decoherence caused by the realistic interaction we consider. 
Since this behavior justifies that the superpositions
of atomic coherent states are relevant with respect to the original 
problem of Schr\"{o}dinger, such a superposition is rightly called 
an atomic Schr\"{o}dinger cat state \cite{benczben997,bencz999,janszky}.
We also note that there are several proposals for the experimental preparation
of these type of states \cite{agarwal997,gg997998}.

Next, in section \ref{catsection} we present the decoherence and dissipation properties 
of  atomic Schr\"{o}dinger cat states, based on numerical computations of their time evolution.
The time scales of decoherence and dissipation differ by orders of magnitude. 
Using this fact, we show how one can make a clear distinction between these two 
processes despite of the interplay between them, 
and we define the characteristic time of decoherence. 
This decoherence time strongly depends on the initial conditions,
notably, it is particularly large for a special 
set of initial cat states \cite{bencz999,bbh999}.
This will be termed as slow decoherence in contrast with the general case which
will be referred to as rapid decoherence.

The interplay between decoherence and energy dissipation 
is the most appreciable in connection with the concept of 
pointer states \cite {zurek981}. It will be shown that when the decoherence is 
rapid, 
then the constituent coherent 
states of the initial  
state are pointer states 
to a very good approximation. However, when there is enough time for 
dissipation, i.e. when decoherence is slow, then the initial atomic coherent 
states themselves evolve into mixtures, and  
therefore a refined scheme of decoherence 
holds.

In order to underline the contrast between 
rapid and slow 
decoherence we superpose four atomic coherent states
corresponding to the vertices of a suitably oriented tetrahedron.
The time evolution of this four component cat state will be studied 
by the aid of the spherical Wigner function in section \ref{wigner}. 
As it is expected, the interaction with the environment 
selects that pair from the initial superposition 
which constitutes a long-lived cat state.


\section{Description of the model}
\label{model}
We consider 
a system of identical two-level atoms interacting 
with the environment of macroscopic number of photon modes. With dipole 
interaction and in the rotating wave approximation
the total system is
described by the following model Hamiltonian:
\begin{equation}
H=\hbar \omega_{\mathrm{a}}J_{z}+ \sum_{k} \hbar \omega _{k}a_{k}^{\dagger
}a_{k}+ \sum_{k} \hbar g_{k}\left( a_{k}^{\dagger }J_{-}+a_{k}J_{+}\right) ,
\label{model_ham}
\end{equation}
where $\omega_{\mathrm{a}}$ is the transition frequency between the two atomic 
energy levels,  $\omega_k$ denote the frequencies of the modes of the 
environment and  $g_k$ are 
coupling constants. $J_+$, $J_-$ and $J_{z}$ are
dimensionless collective atomic operators obeying the usual angular
momentum commutation relations \cite{dicke9}.
This Hamiltonian leads to the interaction picture 
master equation for the reduced density 
matrix of the atomic system, $\rho$ \cite{bsh971a,agarwal974}:
\begin{eqnarray}
\frac{\rm{d}\rho (t) }{\rm{d}t} & = & -\frac{\gamma }{2}
\ (\langle n \rangle+1)\ (J_{+}J_{-}\rho (t)
+\rho (t) J_{+}J_{-}-2J_{-}\rho (t) J_{+})
\nonumber \\
& & -\frac{\gamma }{2}
\ \langle n \rangle\ (J_{-}J_{+}\rho (t)
+\rho (t) J_{-}J_{+}-2J_{+}\rho (t) J_{-}).  \label{mastereq}
\end{eqnarray}
Here $\langle n \rangle$ is the mean number of photons in the environment and
$\gamma $ denotes the damping rate. Note that the same master equation can be obtained 
by considering a low-Q cavity containing Rydberg atoms \cite{haroche985}.

If the state of the atomic system was initially invariant
with respect to the permutations of the atoms, i.e. it was a superposition 
of the totally symmetric Dicke states \cite{dicke9}, the dipole 
interaction described by 
$H_{\mathrm{int}}= \sum_{k} \hbar g_{k}\left( a_{k}^{\dagger }J_{-}+a_{k}J_{+}\right)$ 
in (\ref{model_ham}) would not destroy this symmetry.
Therefore we may restrict
our investigation of the $N$ atom system to the totally symmetric $N+1$
dimensional subspace of the whole Hilbert-space of the atomic system. This subspace 
is isomorphic to 
an angular momentum eigensubspace labeled by $j=N/2$.  
This model has been proven to be valid in
cavity QED experiments with many atoms, as reviewed 
in \cite{haroche985}.

The environment as a static reservoir (represented by the thermal photon modes)  
continuously interacts with the atomic system influencing its dynamics.
As it is obvious, the dissipation of the energy leads to thermal equilibrium in the system,
corresponding to the stationary solution of the master equation (\ref {mastereq}). 
However, as it will be shown here, the same master equation describes also 
a much more interesting process. 
The continuous "monitoring" \cite{zurek981} of the 
atomic system by the environment results in the total loss of the 
coherence of the quantum superpositions in the system.
This decoherence process is generally extremely fast compared to the dissipation, except for special
initial conditions  
which will be discussed in subsection \ref{times}.


\section {The initial stage of the time evolution}
\label{robust}
In this section we apply 
general concepts to our system in order to find the initial states 
for the master equation (\ref {mastereq}) which
are relevant to the original problem of Schr\"{o}dinger \cite{sch}
concerning the unobservability of macroscopic superpositions.

First we consider the short-time behavior of the total
system. The (by assumption pure) system+environment state, $\left| \Psi \right\rangle$
can be written in the Schmidt representation \cite{schmidt,kubzeh973,EkertKnight,decref2} 
at any time as 
\be 
\left| \Psi (t)\right\rangle =\sum_{k}\sqrt{p_{k}(t)}\left| \varphi
_{k}(t)\right\rangle \left| \Phi _{k}(t)\right\rangle
\label {schmidt},
\ee
where $\left| \varphi _{k}\right\rangle $\ and $\left| \Phi
_{k}\right\rangle $\ are elements of certain orthonormal bases (Schmidt bases) 
of the system and the environment, respectively.

At zero temperature the photon field of the present model is  in its pure 
vacuum state $\left| 0\right\rangle$, therefore the initial state factorizes as
\be
\left| \Psi (0)\right\rangle =\left| \varphi _{0}(0)\right\rangle
\left| 0\right\rangle,
\ee
i.e. $p_{0}(0)=1$ and $p_{k}(0)=0$ for $k\neq 0$.  Due to the
interaction this product state  evolves into
a more general Schmidt sum like Eq. (\ref{schmidt}), or in other words it 
turns into an entangled state. 
The time scale of this entanglement or de-separation can be obtained 
\cite {kubzeh973} in leading order of time as
\be
p_{0}(t)=1-At^{2},
\ee with
\be A=\sum_{k\neq 0,l\neq 0}\left| \left\langle \varphi _{k}(0)\right|
\left\langle \Phi _{l}(0)\right| H_{\mathrm{int}}\left| \varphi _{0}(0)\right\rangle
\left| 0\right\rangle \right| ^{2}.
\ee
This quantity can be called  the rate of entanglement. Using the 
explicit form of the interaction Hamiltonian $H_{\mathrm{int}}$,
a straightforward calculation leads to
\be
A={\cal C}\left( J_{+},J_{-}\right) :=\left\langle J_{+}J_{-}\right\rangle
-\left\langle J_{+}\right\rangle \left\langle J_{-}\right\rangle,
\label{normcorr} 
\ee
i.e. in our system the rate of entanglement is the normally
ordered correlation function of the operators $J_{+}$\ and $J_{-}$. 

Let us turn to the 
case of finite temperatures,
when the total system has to be represented by a 
mixed state 
even at $t=0$. The linear entropy, defined as
\be
S_{\mathrm{lin}}={\rm Tr}(\rho - \rho^2), \label {linent}
\ee
can be regarded as a relevant measure of decoherence \cite{decref2,zurek993}.
Restricting ourselves again to the initial regime of the time evolution, 
we can make use of the master equation (\ref {mastereq}) and calculate the 
time derivative of the linear entropy at $t=0$:
\be
\left( \frac{\partial S_{\mathrm{lin}}}{\partial t}\right) _{t=0}=\gamma \left(
\left\langle n\right\rangle {\cal C}\left( J_{-},J_{+}\right) +\left( \left\langle
n\right\rangle +1\right) {\cal C}\left( J_{+},J_{-}\right) \right).
\label{sat0}
\ee
The normally (antinormally) ordered correlation function,
${\cal C}\left( J_{+},J_{-}\right)$ (${\cal C}\left( J_{-},J_{+}\right)$), 
disappears in the eigenstate $\left| j, m=-j \right\rangle$  
($\left|  j, m=j \right\rangle$) of  $J_{-}$\ ($J_{+}$).
However, the collective atomic operators $J_{-}$ and $J_{+}$ have no simultaneous eigenstates 
which would annullate the right hand side of Eq. (\ref {sat0}).
Nevertheless, we are going to show that if the number of atoms $N=j/2$
is large enough, then the 
correlation functions in Eq. (\ref{sat0}) are negligible in a class of 
states called atomic coherent states \cite{acgt972}. 
These states are 
labeled by a complex parameter 
$\tau=\tan(\beta/2)\exp(-i\phi)$ (for the angles $\beta$ and $\phi$ see Fig. \ref{sphere}) 
and can be expanded in terms of 
the eigenstates of the operator $J_z$ (Dicke states) as
\be  |\tau \rangle=\sum^j_{m=-j} \left( 
\begin {array}{cc} 2j \\ j+m \end {array} \right)^{1 \over 2} {\tau^{j+m} \over
{(1+|\tau|^2)^j}} |j,m\rangle \label{cohs}. \ee

For large $j$, the atomic coherent states are approximate
eigenstates of the operators $J_{-}$ and  $J_{+}$ \cite{acgt972,bbh999}.
This statement is understood in the sense that the square of the cosine of 
the angle $\alpha$ between $\left| \tau \right\rangle$ and 
$J_- \left| \tau \right\rangle$:
\be
\cos^2 \alpha={{\left| \left\langle \tau \left| J_- \right| \tau \right\rangle \right|^2}
\over
{\left\langle \tau \left| \right. \tau \right\rangle
\left\langle \tau \left| J_+ J_- \right| \tau \right\rangle}}
\ee
differs from unity by a factor which scales as $(j\tau^2)^{-1}$. 
Thus $\alpha$ becomes negligible in the $j \to \infty$ limit for finite $\tau$ 
\cite {bbh999}. The same 
statement holds for the operator $J_+$, therefore both
correlation functions 
in Eq. (\ref {sat0}) are indeed negligible in the atomic 
coherent states (\ref{cohs}).

This suggests that the atomic
coherent states are rather stable against the decoherence induced by
the photon modes, i.e. they can serve as a model of 
classical-like
macroscopic quantum states. This result is in analogy with the stability of the oscillator
coherent states obtained in \cite{zurek993}.

Two such states, $\left| \tau_1 \right\rangle$ and
$\left| \tau_2 \right\rangle$ can be considered as macroscopically distinct, 
whenever
the distance between the parameters $\tau_1$ and $\tau_2$ is
sufficiently large on the complex plane. This implies that the
coherent superposition of these states, which will be introduced in the
next section, yields an appropriate model of the original 
paradox of Schr\"{o}dinger.


\section {Decoherence of atomic Schr\"{o}dinger cats}
\label{catsection}
  
\begin{figure}[htbp]
\begin{center}
\epsfxsize=3.375in
\epsfbox{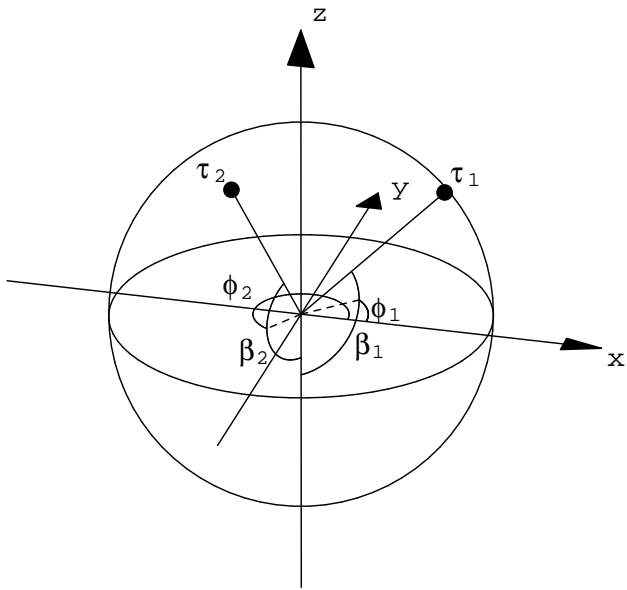}
\end{center}
\caption{
Scheme of an atomic Schr\"{o}dinger cat state defined by Eq. (\ref{cats}).
The points labeled by $\tau_1$ and $\tau_2$ represent the corresponding atomic coherent
states on the surface of the Bloch-sphere. The angles defining the  
$\tau$ parameters are also shown.
}
\label{sphere}
\end{figure}

Based on the results of the previous section,
the superpositions 
\begin{equation}
|\Psi _{12}\rangle ={\frac{{|\tau_1\rangle +|\tau_2\rangle }}
{\sqrt{2(1+{\rm{Re}}\; \langle \tau_1|\tau_2\rangle )}}}
\label{cats}
\end{equation}
will be called atomic Schr\"{o}dinger cat states \cite {benczben997,bencz999,fbencz000},
see Fig. \ref{sphere}. 
Now we are going to present our results on the decoherence and dissipation dynamics of
these type of  
states.
%


\subsection{Time scales}
\label{times}
\begin{figure}[htbp]
\begin{center}
\epsfxsize=3.375in
\epsfbox{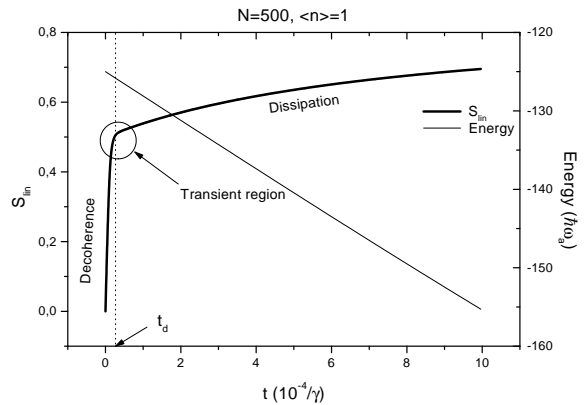}
\end{center}
\caption{The two regimes of the time evolution.
(Initially: $\tau_1=\tan\pi/4$, $\tau_2=0$.)
 The number of atoms is $N=500$ and the average number of photons is $\langle n 
\rangle=1$, $\td \approx 6 \times 10^{-5}/\gamma$}. 
\label{scales}
\end{figure}

A typical result of the  numerical integration of Eq. (\ref {mastereq}) is
that the time evolution of the 
states given by Eq. (\ref{cats}) 
can be characterized by two different time scales, as illustrated by 
Fig. \ref{scales}, where the linear entropy and the energy of the 
atomic system is plotted versus time. 
As we can see, 
there exists a time instant $\td$ (marked with an arrow in Fig. \ref{scales}) 
when the character of the physical process changes radically. 
Initially $S_{\mathrm{lin}}(t)$ increases rapidly 
while 
the dissipated energy of the atoms is just a small  fraction of that part of the energy which 
will  eventually be transferred to the environment.  
On the other hand, for longer times $t\gg \td$  both curves 
change on the same time scale. 
The energy of the atomic system 
decays exponentially as a function
of time allowing for identifying the characteristic time of the dissipation $\tdiss$,
with the inverse of the exponent.
(We note that in Fig. \ref{scales} the plotted time interval is much
shorter than 
$\tdiss$, thus the exponential behavior is not seen.)
More detailed calculations have shown that 
for high temperatures 
energy and linear entropy exhibit similar exponential behavior in the second 
regime of the time evolution. Their exponents coincide with 2-3\% relative error.
This implies that the 
initial stage of the time  
evolution is dominated by decoherence 
while after $\td$ the dissipation determines the dynamics. 
Accordingly we define the characteristic time of the decoherence as the instant
when the slope of the curve $S_{\mathrm{lin}}(t)$  decreases appreciably. We note that 
$\td$ defined in this way is in accordance with the decoherence time defined 
previously in \cite{bencz999} for
a specific initial state.

It is remarkable that although a few hundred atoms do not really constitute a
macroscopic
system, the difference of the time scales is obviously seen in Fig. \ref{scales}.
It is generally true that the larger $N$ is, the more naturally and sharply 
the time evolution splits into two regimes. 

Now we turn to the investigation of the dependence of the decoherence time $\td$
on the initial conditions.  
\begin{figure}[htbp]
\begin{center}
\epsfxsize=3.375in
\epsfbox{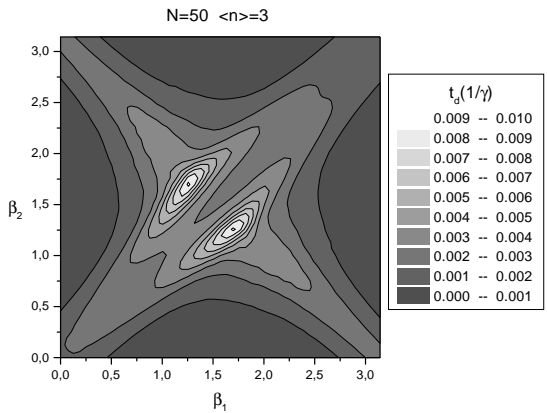}
\end{center}
\caption{The dependence of the characteristic time of the decoherence on
the parameters of the initial Schr\"{o}dinger cat state: $\tau_1=\tan\beta_1/2$, 
$\tau_2=\tan\beta_2/2$. 
The number of atoms is $N=50$ and the average number of photons is $\langle n 
\rangle=3$.}
\label{dectimes}
\end{figure}
Fig. \ref{dectimes}
shows the contour plot of the decoherence time versus the parameters $\beta_1$ 
and $\beta_2$ (see Fig. \ref{sphere}) of the initial atomic Schr\"{o}dinger cat 
state (\ref{cats}). We have set $\phi_1=\phi_2=0$ for simplicity.
As we can see, the effect of decoherence is remarkably slower when $\beta_1\approx
\beta_2$ which was expected since 
in this case the overlap of the two initial coherent states is not negligible,
so these states can not be considered as ``macroscopically distinct''.
Much more surprising is the fact that cat states which were initially 
symmetric with respect to the 
$(x,y)$ plane
(i.~e.~$\beta_1\approx\pi-\beta_2$ ) also decohere slower 
\cite{bencz999,fbencz000}, but it is in accordance with the analytical estimations of Braun
et.~al.~\cite{bbh999} 
In the following sections we shall refer to these states as {\it symmetric}
ones.


\subsection {The direction of the decoherence}
\label {pointer}
We saw in the previous subsection that the interplay between decoherence
and dissipation is reflected in the time evolution of the 
superpositions given by
Eq. (\ref {cats}). In this subsection we shall focus on 
the direction of the process resulting from the dynamics governed by the master
equation (\ref {mastereq}).

According to the  general theory of environment induced decoherence 
\cite{zurek981,decref1,decref2,zurek993,venug000}, 
the interaction with a large number of degrees of freedom 
selects naturally 
the so-called pointer basis \cite{zurek981} in the Hilbert-space of the system subject 
to decoherence.
This process is claimed to favor the constituent states of the 
pointer basis in the sense that 
the system  is driven towards a classical statistical mixture of these states. 
Thus, from the present point of view $\rho(\td)$ is the relevant quantity to be examined.

Recalling the analytical results of sec. \ref {robust}, it seems plausible to expect 
that the atomic coherent states (\ref {cohs}) will be pointer states.

By introducing the 
density matrix which corresponds to the classical statistical mixture of 
the initial coherent states:
\be
\rho_{\mathrm{cl}}(\tau_1,\tau_2)={1 \over 2} \left( |\tau_1 \rangle 
\langle \tau_1| +
|\tau_2\rangle  \langle \tau_2| \right),
\label{rhocl}
\ee  
the expected scheme of the decoherence reads:
\be
|\Psi _{12}\rangle \langle \Psi _{12}| \rightarrow \rho_{\mathrm{cl}}(\tau_1,\tau_2).
\label{scheme}
\ee
We shall refer to $\rho_{\mathrm{cl}}$ as the classical density matrix. 

\begin{figure}[htbp]
\begin{center}
\epsfxsize=3.375in
\epsfbox{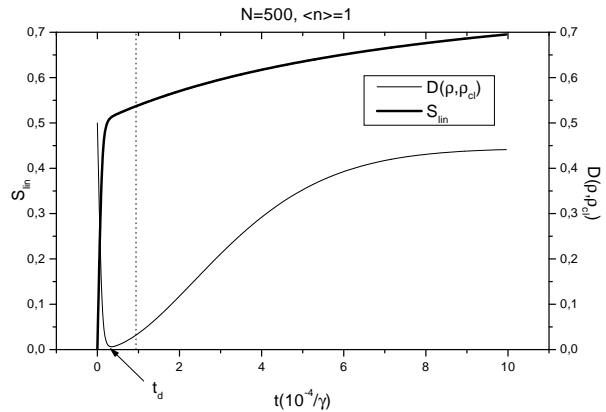}
\end{center}
\caption{The linear entropy $S_{{\mathrm{lin}}}$ and the distance $D$ between $\rho$ and
$\rho_{\mathrm{cl}}$ (defined by Eq. \protect{(\ref {distance}})) in the case of a rapidly decohering 
Schr\"{o}dinger cat state ($\tau_1=\tan\pi/4$, $\tau_2=0$). 
The number of atoms is $N=500$ and the average number of photons is $\langle n 
\rangle=1$.}
\label{fastdec}
\end{figure}
The distance between the 
actual density matrix $\rho(t)$
and $\rho_{\mathrm{cl}}$, defined with
\be
D \left(\rho(t),\rho_{\mathrm{cl}} \right)={\rm Tr}\left[ \left( \rho(t)-\rho_{\mathrm{cl}} \right)^2 \right],
\label{distance}
\ee
is always decreasing fast.
Except for the case of slowly decohering cat states which will be discussed below,  
$D \left(\rho(t),\rho_{\mathrm{cl}} \right)$ reaches its minimal value 
at the {\it decoherence time},  see  Fig. \ref {fastdec}.
This minimal value is very close to zero implying that the 
density matrix of the system at this instant is nearly the same as the classical density
matrix (\ref{rhocl}).
This fact 
justifies the definition of the characteristic time of the
decoherence in section \ref {times}, 
and it is in excellent agreement with the decoherence
scheme (\ref{scheme}). 

Due to the exceptionally slow decoherence, we have to modify 
this picture if the initial  state is a symmetric superposition. 
In this case the decoherence time is so long that the atomic coherent 
constituents of the initial state are also appreciably affected 
by the time evolution until the decoherence time, $t_d$.  
The state of the atomic system at  $t_d$ will be a mixture, 
which is the same as if the system had started from  
$\rho_{\mathrm{cl}}$ at t=0. In other words the evolution 
follows the modified scheme: 
\be
|\Psi _{12}\rangle \langle \Psi _{12}| \rightarrow 
\tilde{\rho}_{\mathrm{cl}}(\tau_1, \tau_2, t)
\label{scheme2}
\ee
where the time dependent classical density matrix
$\tilde{\rho}_{\mathrm{cl}}(\tau_1, \tau_2, t)$ is the one 
which would evolve from the statistical mixture
(\ref{rhocl}) 
$\rho_{\mathrm{cl}}(\tau_1,\tau_2)=\tilde{\rho}_{\mathrm{cl}}(\tau_1, \tau_2, t=0)$ 
according to the same master equation (\ref {mastereq})
as the actual atomic density matrix. 
The distance between the time dependent classical density matrix, $\tilde{\rho}_{\mathrm{cl}}$ 
and  $\rho(t)$ becomes negligible at $\td$, and asymptotically reaches zero for long times
in the case of all the initial conditions. 

\section {Wigner functions of four component Schr\"{o}dinger cat states}
\label {wigner}
\begin{figure}[htbp]
\begin{center}
\epsfxsize=3.375in
\epsfbox{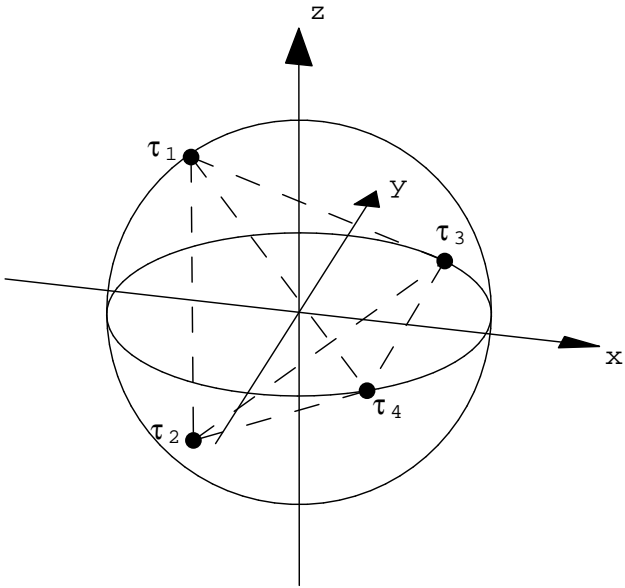}
\end{center}
\caption{Phase space scheme of the 4 component cat state. The atomic coherent states 
constituting the superposition (\ref{4cats}) are represented by the points labeled by
$\tau_1,\ \ldots,\ \tau_4$.
They are arranged to form the vertices of a tetrahedron
as shown.}
\label{tetrahedron}
\end{figure}

The results of the previous section have shown that both 
the characteristic time and the direction of the decoherence
strongly depend on the initial conditions. 
In this section we illustrate this fact by tracking
the decoherence of the superposition of four atomic coherent states: 
\be
\left| \Psi_{1234} \right \rangle = {\frac{{|\tau_1\rangle +|\tau_2\rangle+|\tau_3\rangle+|\tau_4\rangle }}
{\sqrt{2(2+{\rm{Re}}\; \sum_{i>k}\langle \tau_i|\tau_k\rangle )}}}. 
\label{4cats}
\ee
Since four points on the surface of a sphere are not distinguished with respect to each other
if and only if they are the vertices of a regular 
tetrahedron inscribed in the sphere, we set the
components of $\left| \Psi_{1234} \right \rangle$ according to this pattern.
On the other hand,
the $z$ axis is distinguished in the present model because of the form of the
Hamilton operator (\ref{model_ham}), therefore we orient the tetrahedron 
with one edge 
parallel to the $z$ axis and the opposite edge parallel to the $y$ axis, 
see Fig. \ref{tetrahedron}.
Although we have in principle 
two 
substantially different ways 
of considering the state represented by Fig. \ref{tetrahedron} as the superposition of
{\it two} atomic Schr\"odinger cat states,
according to the results of the previous section
one expects that the environment naturally selects one of these possibilities via the 
different time evolutions:
the quantum coherence between the {components} of the symmetric pair
 $\left| \Psi_{12} \right \rangle \propto \left| \tau_{1} \right \rangle + \left| \tau_{2} \right \rangle$    
disappears slowly, 
while all the other pairs are rapidly decohering superpositions.

We are going to
visualize the decoherence process of $\left| \Psi_{1234} \right \rangle$ 
by the aid of the spherical Wigner function. 
It is a real function over the unit sphere (which is the appropriate phase space in the present case)
being in a linear one-to-one correspondence with the density matrix of 
the atomic system. 

Following the 
construction of  Agarwal \cite{agarwal981}, the Wigner function 
corresponding to a reduced density matrix $\rho$ of the atomic system
is defined as the expectation value
of the operator kernel $\Delta(\theta,\phi)$:
\be
W(\theta,\phi)={\rm Tr}\left( \rho \Delta(\theta,\phi) \right),
\label{wfunct}
\ee   
with
\be
\Delta(\theta,\phi)=\sum_{K=0}^N \sum_{Q=-K}^K T^\dagger_{KQ} Y_{KQ}(\theta,\phi).
\ee
Here the $T_{KQ}$ are the multipole operators and $Y_{KQ}$ denote 
the spherical harmonics \cite{BL}. For previous applications of the spherical
Wigner function see \cite{benczben997,bencz999,DAS,CB96,Brif,Chumakov99}.

\begin{figure}[htbp]
\begin{center}
\epsfxsize=6.5in
\epsfbox{
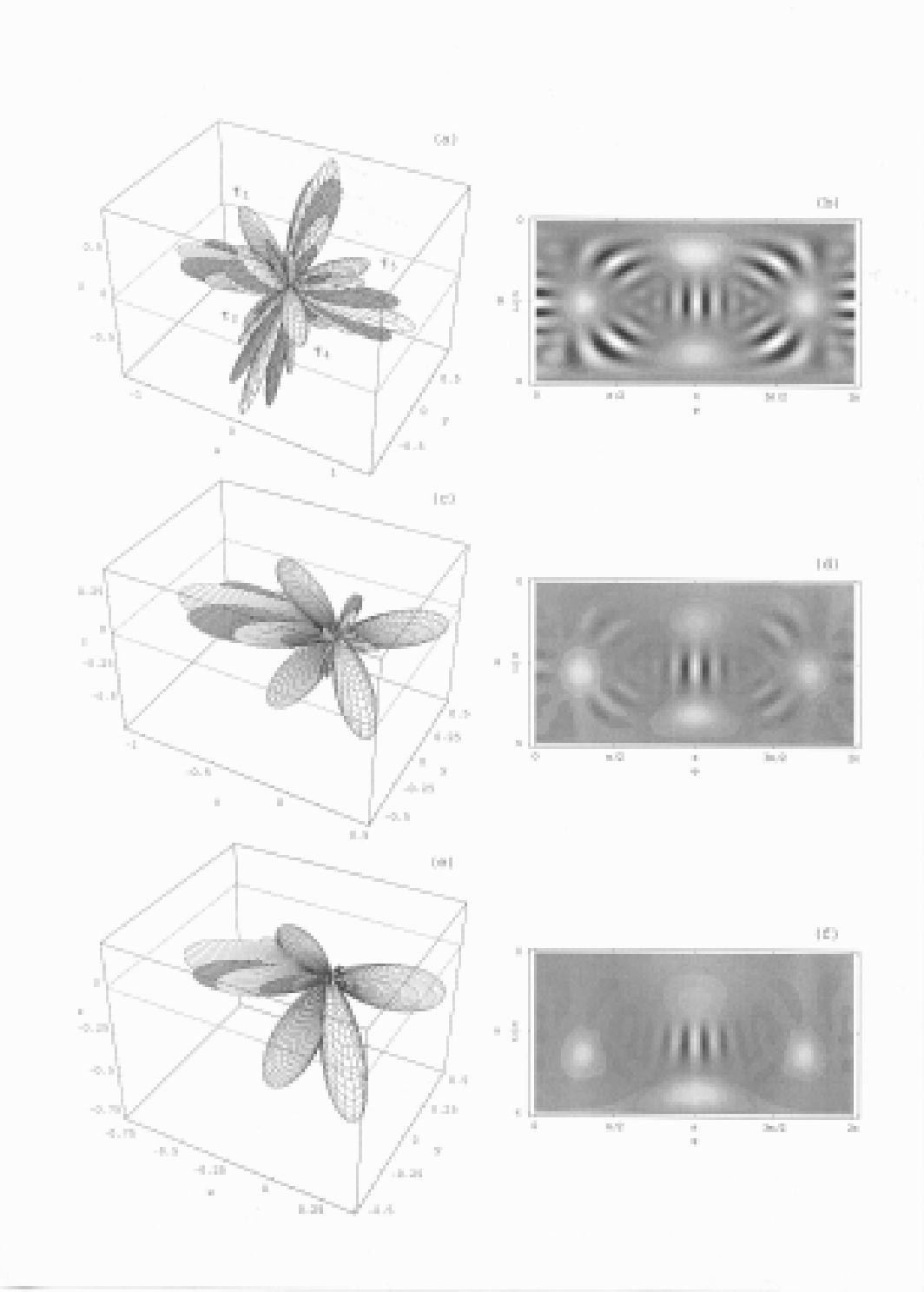
}
\end{center}
\caption{Wigner view of the decoherence of the 4 component cat state.
We plot the spherical Wigner function (\ref{wfunct}) both as a polar
plot (a, c, e) and as a contour plot (b, d, f). A polar plot is obtained
by measuring the absolute value of the function in the corresponding
direction, 
and the resulting surface is shown in light where the 
values of the Wigner function are positive, and
in dark where they are negative. Similarly, light shades of the contour
plot correspond to positive function values, while dark shades mean 
negative function values. Plots (a, b) show the initial state, i.e.
the superpositon $\left| \Psi_{1234} \right \rangle$
 (Eq. (\ref{4cats})) as shown in  Fig. \ref{tetrahedron}. In 
(a) we also label the lobes corresponding
to the initial 
coherent constituents. Plots (c, d) show the spherical Wigner function
at $t=0.015/\gamma$, while plots (e, f) correspond to 
$t=0.04/\gamma$.
}
\label{wigfig}
\end{figure}

The Wigner function  (\ref{wfunct}) suggestively
maps the time evolution of the state (\ref {4cats}) 
onto the unit sphere, as shown 
in Fig. \ref{wigfig}. Here we present the state
of the atomic system at three time instants,  and we plot the corresponding 
Wigner functions both as a polar plot and as a contour plot. Dark shades mean negative,
light shades mean positive function values.
The four positive lobes, pointing from the center to the vertices of the tetrahedron shown 
in Fig. \ref{tetrahedron},  
correspond  to the four atomic coherent states in (\ref {4cats}). 
Due to the dissipation all these lobes will move slowly downwards.
The initial interference pattern (Figs. \ref{wigfig}(a) and (b)) has the regularity of 
the 
tetrahedron, there are equally pronounced oscillations along all the edges, representing the quantum 
coherence between the coherent states.

Figs. \ref{wigfig}(c) and (d) depict the situation after a time  
which is short in the sense that the 
the shapes of the lobes of the coherent states are not appreciably affected 
(no dissipation),  but the interference 
is already negligible between them, except for the single pair along the vertical edge of 
the tetrahedron. As it is seen from Fig. \ref{tetrahedron} this is the pair which 
represented initially the {\it symmetric} atomic Schr\"odinger cat state
$\propto\left| \tau_{1} \right \rangle + \left| \tau_{2} \right \rangle$ in (\ref {4cats}).
The coherence between 
the components of this pair of states  is nearly unaffected as shown by the strong oscillations. 

A qualitatively different stage of the time evolution 
is shown in Figs. \ref{wigfig}(e) and (f) at a later time. 
The coherent constituents are already affected by the dissipation, 
the uppermost one rather strongly,
but the quantum coherence between the components of the symmetric pair is still present. 
On the contrary, the interference between all the other components has already disappeared.

\onecolumn

\twocolumn

In view of the present results, if the initial state of the
atomic system is a superposition of coherent states so that there are
symmetric pairs of coherent states in the expansion, 
then the coherence between the components of these
symmetric pairs will survive much longer than between any other terms.


\section{Conclusions}

We have investigated 
the decoherence of superpositions of macroscopically distinct quantum states in a
system of two-level atoms embedded in the environment of thermal photon modes.
Utilizing the Schmidt decomposition and the linear entropy, we have shown that atomic
coherent states are robust against decoherence, both at zero
and non-zero temperatures. This result is in analogy with the harmonic oscillator
case and justifies the definition of atomic 
Schr\"{o}dinger cat states as superpositions of atomic coherent states.

By solving the master equation (\ref{mastereq}) we have identified 
two different regimes of the time evolution with the help of the linear entropy. The first one is
dominated by decoherence while the second one is governed by dissipation.
Based on several computational runs focusing on the characteristic times it was found 
that $\td$ decreases much faster than $\tdiss$ as the function of the number 
of atoms, $N$. Consequently $\td$ becomes many orders of magnitude smaller than the 
characteristic time of dissipation for macroscopical samples, and 
even for e.g. $N=500$ atoms and an average photon number 
$\langle n \rangle = 1$ the ratio $\tdiss/\td$ is around a few hundred depending 
on the initial conditions. 
However, there are very important exceptional cases, called slow decoherence, 
when the atomic coherent 
states constituting the initial atomic Schr\"{o}dinger cat state are symmetric with
respect to the equator of the Bloch sphere. 

Using a new measure $D$, we have shown that at the characteristic time of decoherence the  system is always
very close to the state described by the time dependent classical density matrix. Apart from
the exceptional case of 
slow decoherence, the coherent states appearing in (\ref {cats}) 
are approximate pointer states. When due to its symmetry the initial cat state is a 
long-lived superposition, also its constituent coherent states have time to transform 
into mixtures until $\td$. We have given a modified scheme of decoherence which is 
valid also for slow decoherence. Accordingly, the orthonormal pointer basis consists of the eigenstates of 
$\tilde{\rho}_{\mathrm{cl}}(\tau_1, \tau_2, \td)$.

We have demonstrated the important difference between rapid and slow  decoherence by
tracking the time evolution of a four component superposition with the help
of the spherical Wigner function. The initial interference pattern having the
symmetry of a tetrahedron rapidly disappears except for the single slowly decohering
pair.


\section*{Acknowledgments}

The authors thank F.~Haake and J.~Janszky for fruitful discussions.
The present work was supported by the Hungarian Scientific Research Fund 
(OTKA) under contracts T022281, T032920, F023336 and M028418.


\end{document}